\documentclass[aps,prb,showpacs,preprint,amsfonts,amssymb,amsmath]{revtex4}

\usepackage{graphicx}
\begin{document}

\title{Heat generation by electric current in mesoscopic devices}

\author{Qing-feng Sun$^{1}$ and X. C. Xie$^{1,2}$}
\affiliation{ $^1$Beijing National Lab for Condensed Matter
Physics and Institute of Physics, Chinese Academy of Sciences,
Beijing 100080,
China;\\
$^2$Department of Physics, Oklahoma State University, Stillwater,
Oklahoma 74078
 }

\date{\today}

\begin{abstract}
We study the heat generation in a nano-device with an electric
current passing through the device. For the first time, a general
formula for the heat generation is derived by using the
nonequilibrium Keldysh Green functions. This formula can be
applied in both the linear and nonlinear transport regions, for
time-dependent systems, and with multi-terminal devices. The
formula is also valid when the nano-device contains various
interactions. As an application of the formula, the heat
generation of a lead-quantum dot-lead system is investigated. The
dc and ac biases are studied in detail. We find several
interesting behaviors that are unique to nanostructures, revealing
significant difference from heat generation in macroscopic
systems.
\end{abstract}


\maketitle

In the past two decades, with the development of micro machining
technology, the operational speed and integrating techniques of
semiconducting electronics have made enormous progress. At the
present time, the size of the electron devices has reached about
$90nm$ in mainstream CPU, and its operation frequency has reached
several GHz. With these advancements, one critical issue, the
dissipation of the conducting electrons (i.e. the heat generation
from electric currents), has emerged. This dissipation in
nano-devices strongly hinders further development of the
semiconductor electronics. Thus, it is important to uncover the
laws of heat generation induced by electric currents in
nano-devices. From a microscopic point of view, the main heat
generation is due to electron-phonon interactions, through which
the energy associated with the electric current in an electronic
system is transferred to the phonon system in the form of heat.
The heat generation in macroscopic systems, the Joule effect, is
well known. However, similar studies in nanostructures are
lacking.

As far as transport is concerned, many studies have been carried
out to investigate the electronic conduction through nano-devices
in the presence of electron-phonon interactions. For example,
recent experiments have observed the extra conductance peaks in a
single-$C_{60}$ transistor\cite{ref1} and in suspended carbon
nanotubes\cite{ref2,ref3}. These extra peaks are due to a phonon
being either absorbed or emitted while electrons are interacting
with the vibrational or breathing phonon modes. On the theoretical
front, the quantum-dot (QD) systems containing electron-phonon
interactions have been extensively
investigated\cite{ref4,ref5,ref6}. These studies concentrate on
physical properties such as the current, the conductance, the
noise of the current, and the spectral function, {\it etc}. Thus,
the previous work have mainly focused on the electronic transport
properties and neglected the heat generation due to electric
current.

In this Letter, we investigate the heat generation while a current
flows through a nano-device. For the first time, a general exact
formula [Eq.(6) or Eq.(8)] of the heat generation is derived. It
expresses the heat generation in terms of the two-particle Green
functions of the system, as conventionally done for the
conductance formula. This formula is applicable for either linear
or nonlinear transport, and also for time-dependent systems. The
condition for zero heat generation is readily achieved with this
formula. To illustrate the usefulness of the formalism, we
investigate the physical properties of heat generation in a single
QD system. With a dc bias and at zero temperature, a threshold
bias $V_t$ for non-zero heat generation is found. With an ac bias,
a finite heat generation is found even at the time when the
current vanishes. These behaviors are quite different from our
knowledge of the heat generation in macroscopic systems.

Consider the device as a mesoscopic center region coupled to
electro-leads that act as electron reservoirs (see Fig.1a). In the
central region, an electron-phonon interaction exists, so the heat
(i.e. the phonon) is generated when the current flows through the
device. This system can be described by the following Hamiltonian:
\begin{equation}
  H = \sum\limits_q \omega_q \hat{a}_q^{\dagger} \hat{a}_q
  + \sum\limits_q \lambda_q (\hat{a}_q \hat{A}_q^{\dagger} +\hat{a}_q^{\dagger} \hat{A}_q)
  + H_e.
\end{equation}
Here the first term is the Hamiltonian of the free phonon in the
central region, with $\hat{a}_q(\hat{a}_q^{\dagger})$ being the
phonon annihilation (creation) operator. The second term expresses
the electron-phonon interaction and the operator $\hat{A}_q =
\sum_{i,j} M_{q,ij}\hat{d}_i^{\dagger} \hat{d}_j$, in which
$\hat{d}_i$ is the electron annihilation operator for the state
$i$ in the central region. The last term $H_e$ describes the
electron system and is of the form:
\begin{eqnarray}
  H_e & = & \sum\limits_i \epsilon_i(t) \hat{d}_i^{\dagger}
  \hat{d}_i
  +H_{int} + \sum\limits_{\alpha, k} \epsilon_{\alpha k}(t)
   \hat{c}_{\alpha k}^{\dagger} \hat{c}_{\alpha k} \nonumber\\
   & + & \sum\limits_{\alpha,k,i} [ t_{\alpha k i} \hat{c}_{\alpha
   k}^{\dagger} \hat{d}_i +H.c],
\end{eqnarray}
where $i$ is the index of the state in the central region, which
includes both the  energy level and spin indices. The center
scattering region can be a QD, double QDs, a molecule, a nanotube,
{\it etc}. Moreover, the center region can contain various
interactions $H_{int}$, e.g. the electron-electron Coulomb
interaction $\sum_{i,j(i\not=j)} U_{ij} \hat{d}_i^{\dagger}
\hat{d}_i \hat{d}_j^{\dagger} \hat{d}_j$, the hopping term
$\sum_{i,j(i\not=j)} t_{ij} \hat{d}_i^{\dagger} \hat{d}_j$, {\it
etc}. The third and forth terms in Eq.(2) are for the leads and
the coupling between the leads and the central region,
respectively, in which $\alpha$ is the lead index. In Eq.(2), the
leads are the normal metal or semiconductor. In fact, the derived
heat generation formula given below is also valid for
superconducting and ferromagnetic leads. The device may be under a
high-frequency ac bias, a time-dependent gate voltage, or a
microwave radiation, in those cases, the single particle energies
$\epsilon_i$, $\epsilon_{\alpha k}$, and the hopping element
$t_{\alpha k i}$ are all time-dependent.\cite{ref7}

Assume that the isolated phonon system acts as a phonon reservoir
and that the phonons are in the equilibrium Boson distribution
$N_q =1/[\exp(\omega_q /k_B T_{ph})-1]$ with the phonon
temperature $T_{ph}$.\cite{note1} Then the heat generation $Q(t)$
at the time $t$, i.e. the energy flow from the electrons to the
phonons, can be calculated from the time-evolution of the operator
$\hat{E}_{ph} =\sum_q \omega_q \hat{a}_q^{\dagger}(t)
\hat{a}_q(t)$:
\begin{equation}
 Q(t) =  \langle d\hat{E}_{ph}/dt \rangle
  = -2 {\bf Re} \sum_q  \omega_q \lambda_q  i
  \langle \hat{a}_q^{\dagger}(t) \hat{A}_q(t) \rangle.
\end{equation}
We introduce the standard Keldysh nonequilibrium Green functions
in the derivation: $\langle\langle X(\tau)|Y(\tau')
\rangle\rangle^c \equiv -i \langle T_c [X(\tau)Y(\tau')] \rangle$,
$\langle\langle X(t)|Y(t') \rangle\rangle^< \equiv -i \langle
Y(t')X(t) \rangle$, and $\langle\langle X(t)|Y(t')
\rangle\rangle^r \equiv -i \theta(t-t')\langle [X(t),Y(t')]
\rangle$. Here $X$ and $Y$ represent the Boson operators, e.g.
$\hat{a}_q$, $\hat{A}_q^{\dagger}$, $\hat{d}_i^{\dagger}
\hat{d}_j$, {\it etc.}, the times $\tau$ and $\tau'$ are in the
complex-time contour and $t$, $t'$ are the usual
time,\cite{ref7,ref9,ref10} and $T_c$ is the contour-ordering
operator. By using these Green functions, the heat generation of
Eq.(3) can be expressed as:
\begin{eqnarray}
 Q(t) =  2 {\bf Re} \sum\limits_q  \omega_q \lambda_q
  \langle\langle \hat{A}_q(t)| \hat{a}_q^{\dagger}(t)
  \rangle\rangle^<.
\end{eqnarray}
Since the isolated phonon Hamiltonian has no interaction, one
obtains the following equations by using the Wick's
theorem:\cite{ref7,ref9,ref10}
\begin{eqnarray}
 \langle\langle
 \hat{A}_q(\tau)|\hat{a}_q^{\dagger}(\tau')\rangle\rangle^c
   =\lambda_q \int_c d\tau_1
   \langle\langle
 \hat{A}_q(\tau)|\hat{A}_q^{\dagger}(\tau_1)\rangle\rangle^c
  g_q^{c}(\tau_1,\tau') \nonumber,
\end{eqnarray}
where $g_q^c(\tau,\tau') =   \langle\langle
 \hat{a}_q(\tau)|\hat{a}_q^{\dagger}(\tau')\rangle\rangle^c_0$ is
the free Green function of the decoupled phonon system (i.e. when
$\lambda_q =0$). After the Green function on the complex contour
is solved, the lesser component is simply of the form:
\begin{eqnarray}
 \langle\langle
 \hat{A}_q(t)|\hat{a}_q^{\dagger}(t')\rangle\rangle^<
   =\lambda_q \int d t_1
 \left\{   \langle\langle
 \hat{A}_q(t)|\hat{A}_q^{\dagger}(t_1)\rangle\rangle^r
  g_q^{<}(t_1,t')
  +
     \langle\langle
 \hat{A}_q(t)|\hat{A}_q^{\dagger}(t_1)\rangle\rangle^<
  g_q^{a}(t_1,t') \right\}.
\end{eqnarray}
Here the free phonon Green functions $g_q^{<}(t,t') = -i N_q e^{-i
\omega_q (t-t')}$ and $g_q^a(t,t') = i\theta(-t+t')e^{-i \omega_q
(t-t')}$. Substituting $\langle\langle
 \hat{A}_q(t)|\hat{a}_q^{\dagger}(t)\rangle\rangle^<$ into
Eq.(4), the heat generation $Q(t)$ is obtained as:
\begin{eqnarray}
 Q(t) = -2 {\bf Im} \sum\limits_q  \omega_q \lambda_q^2
  \int_{-\infty}^t dt_1 e^{-i\omega_q(t_1-t)}
  \left\{ -N_q \langle\langle
 \hat{A}_q(t)|\hat{A}_q^{\dagger}(t_1)\rangle\rangle^r
  +
     \langle\langle
 \hat{A}_q(t)|\hat{A}_q^{\dagger}(t_1)\rangle\rangle^<
 \right\}.
\end{eqnarray}
If the phonon spectrum is continuous and multi-modes, the
summation $\sum_q$ in Eq.(6) changes into an integral
$\sum_{\beta}\int_0^{\infty}\rho_{\beta}(\omega_q) d \omega_q$,
with the phonon mode index $\beta$ and the phonon density of
states $\rho_{\beta}$. In this case, the heat generation $Q(t)$
is:
\begin{eqnarray}
 Q(t) = -2 {\bf Im} \sum\limits_{\beta} \int_0^{\infty} d\omega_q \rho_{\beta}
  \omega_q \lambda_{\beta q}^2
  \int_{-\infty}^t dt_1 e^{-i\omega_q(t_1-t)}
  \left\{
 \langle\langle
 \hat{A}_{\beta q}(t)|\hat{A}_{\beta q}^{\dagger}(t_1)\rangle\rangle^<
   -N_q \langle\langle
 \hat{A}_{\beta q}(t)|\hat{A}_{\beta q}^{\dagger}(t_1)\rangle\rangle^r
 \right\}.
\end{eqnarray}

Eq.(6) or Eq.(7) is one of the main results of this paper. Without
any approximation, it expresses the heat generation in terms of
the electronic two-particle Green function for the central region,
as commonly done for transport coefficients. This formulation can
be applied to the nonlinear case with a finite bias, to an ac bias
case. It is also valid for various interactions possibly existing
in the central region, or with superconducting or ferromagnetic
leads, {\it etc}.

Next, we consider the steady state case when the Hamiltonian (1)
is time-independent. In this case, the two-particle Green
functions in Eq.(6) depend only on the time difference $t-t_1$. By
introducing the Fourier transform of the Green functions:
$\langle\langle
\hat{A}_q(t)|\hat{A}_q^{\dagger}(t_1)\rangle\rangle^{r,<} = \int
\frac{d\omega}{2\pi} e^{-i\omega(t-t_1)} \langle\langle
\hat{A}_q|\hat{A}_q^{\dagger} \rangle\rangle^{r,<}_{\omega}$, then
Eq.(6) reduces to:
\begin{eqnarray}
 Q = \sum\limits_q  \omega_q \lambda_q^2
  \left[ 2N_q {\bf Im} \langle\langle
 \hat{A}_q|\hat{A}_q^{\dagger}\rangle\rangle^r_{\omega_q}
  +i \langle\langle
  \hat{A}_q|\hat{A}_q^{\dagger}\rangle\rangle^<_{\omega_q}
 \right].
\end{eqnarray}
If a Hamiltonian $H$ has the property that $[\hat{A}_q, H]=0$,
then $\hat{A}_q(t)= \hat{A}_q(0)$ and $\langle\langle
\hat{A}_q|\hat{A}_q^{\dagger}\rangle\rangle^{r,<}_{\omega_q} = 0$
for any non-zero $\omega_q$. Thus, the heat generation is
identically zero from Eq.(8). This means that the value
$[\hat{A}_q, H]$ represents the strength of the dissipation in the
central region. Notice that although the heat generation is zero
for $[\hat{A}_q, H]=0$, the electron spectral function as well as
the electron transport properties are still affected, sometimes
even strongly affected, by the presence of electron-phonon
interactions.\cite{ref10}

If the interactions in the central region are weak, then the
two-particle Green functions in Eq.(8) can be analyzed by making
the approximation $\langle T_c [ \hat{d}_i^{\dagger}(\tau)
\hat{d}_j(\tau) \hat{d}_k^{\dagger}(0) \hat{d}_l(0)] \rangle =
 -
\langle T_c [ \hat{d}_i^{\dagger}(\tau) \hat{d}_l(0) ]\rangle
\langle T_c [\hat{d}_k^{\dagger}(0) \hat{d}_j(\tau) ]
\rangle$.\cite{note2} Under this approximation, $\langle\langle
\hat{A}_q(\tau)|\hat{A}_q^{\dagger}(0)\rangle\rangle^c = -i
\sum_{i,j,k,l}M_{q,ij}M_{q,lk}^* G_{jk}^c(\tau) G_{li}^c(-\tau)$,
where $G_{ij}^c(\tau) = \langle\langle
\hat{d}_i(\tau)|\hat{d}_j^{\dagger}(0)\rangle\rangle^c$ being the
single-electron Green function in the central region. Then the
lesser and retarded components can be easily obtained by using: $
[G^c(\tau)G^c(-\tau)]^<  = G^<(t) G^>(-t)$ and $
[G^c(\tau)G^c(-\tau)]^r =  -i [G^>(t) G^a (-t) +   G^r(t) G^>
(-t)] $ Taking the Fourier transformations and substituting them
into Eq.(8), we have:
\begin{eqnarray}
  Q & = & {\bf Re} \sum\limits_{q,i,j,k,l} \omega_q
  \lambda_q^2 M_{q,ij} M_{q,lk}^* \int \frac{d\omega}{2\pi}
  \left\{ G^<_{jk}(\omega)G^>_{li}(\bar{\omega}) \right. \nonumber\\
  & - & \left. 2N_q \left[ G^>_{jk}(\omega)G^a_{li}(\bar{\omega})
    +  G^r_{jk}(\omega)G^>_{li}(\bar{\omega}) \right]
   \right\},
\end{eqnarray}
where $\bar{\omega}\equiv \omega-\omega_q$. In this expression,
the heat generation is described by the single-electron Green
functions in the central scattering region. In fact, since these
single-electron Green functions have been obtained in previous
transport studies,\cite{ref4,ref5,ref6} so the heat generation can
be calculated straightforwardly by using this expression.

In order to illustrate power of our formalism, we now study two
specific examples. Consider a lead-QD-lead system where the QD has
an electronic level that is coupled to a local single-phonon mode
(e.g. the optical phonon\cite{ref3}, or the vibrational phonon
mode\cite{ref1,ref2}). The Hamiltonian of this device is:
\begin{eqnarray}
   H &= &\omega_q \hat{a}_q^{\dagger}\hat{a}_q +\lambda_q (\hat{a}_q +
   \hat{a}_q^{\dagger}) \hat{d}^{\dagger} \hat{d}
   +\epsilon_d   \hat{d}^{\dagger} \hat{d} \nonumber \\
   & + & \sum\limits_{\alpha,k} \epsilon_{\alpha k}(t)
  \hat{a}_{\alpha k }^{\dagger} \hat{a}_{\alpha k}
  +\sum\limits_{\alpha, k} t_{\alpha}
  (\hat{c}^{\dagger}_{\alpha k}\hat{d}
  +H.c).
\end{eqnarray}
Here $\alpha=L,R$ represent the left and the right leads. For this
Hamiltonian, the operator $\hat{A}_q = \hat{d}^{\dagger} \hat{d}$.
We consider two specific cases: (i) A dc bias is applied to the
left and right leads, with the time-independent $\epsilon_{\alpha
k}$. (ii) An ac bias (without a dc component) is applied to the
two leads, with $\epsilon_{\alpha k}(t) = \epsilon_{\alpha k}^0
\pm \Delta \cos \omega_{ac} t $. In order to solve for the Green
functions, we first make a canonical transformation with the
unitary operator $U = exp\{
(\lambda_q/\omega_q)(\hat{a}^{\dagger}_q-
\hat{a}_q)\hat{d}^{\dagger} \hat{d} \}$. After this
transformation, the Hamiltonian becomes:
\begin{eqnarray}
   H & = & \omega_q \hat{a}_q^{\dagger} \hat{a}_q
 + \tilde{\epsilon}_d  \hat{d}^{\dagger} \hat{d} \nonumber\\
  & + & \sum\limits_{\alpha,k} \epsilon_{\alpha k}(t)
  \hat{a}_{\alpha k }^{\dagger} \hat{a}_{\alpha k}
  +  \sum\limits_{\alpha, k} t_{\alpha}
  (\hat{c}^{\dagger}_{\alpha k}\hat{d}\hat{X}
  +H.c),
\end{eqnarray}
where $\tilde{\epsilon}_d=\epsilon_d-\lambda_q^2/\omega_q$ and
$\hat{X} = exp\{ -(\lambda_q/\omega_q)(\hat{a}^{\dagger}_q-
\hat{a}_q) \}$. Under the canonical transformation, the operator
$\hat{A}_q$ is unchange, i.e. $U\hat{A}_q(t) U^{\dagger} =
U\hat{d}^{\dagger}(t) \hat{d}(t) U^{\dagger} =
\hat{d}^{\dagger}(t) \hat{d}(t) = \hat{A}_q(t)$. Next, we make an
approximation, replacing the operator $\hat{X}$ in the Hamiltonian
(11) by its expectation values $<\hat{X}> = exp\{
-(\lambda_q/\omega_q)^2 (N_q +1/2)\}$. This approximation is valid
when $t_{\alpha} \ll \lambda_q$.\cite{ref6,ref10,ref13} Under this
approximation, the electron-phonon interaction is decoupled,
allowing the two-particle Green functions in the heat generation
formula (6) or (8) to be expressed by the single-particle Green
functions, thus, the problem is solved in a straightforward way.

{\sl (i) The dc bias case}. In this case, the Hamiltonian is
time-independent. After the above approximation, the heat
generation formulation (8) is reduced to:
\begin{eqnarray}
  Q & =& \omega_q \lambda_q^2 \int \frac{d\omega}{2\pi} \left\{
    \tilde{G}_d^<(\omega)
    \tilde{G}^>_d(\bar{\omega}) \right. \nonumber\\
   & - & \left. 2 N_q {\bf Re} \left[ \tilde{G}_d^>(\omega)
    \tilde{G}^a_d(\bar{\omega}) +
    \tilde{G}_d^r(\omega)
    \tilde{G}^>_d(\bar{\omega}) \right] \right\}.
\end{eqnarray}
Here $\tilde{G}_d$ are the QD's single-electron Green functions of
the Hamiltonian (11), and they are easily obtained as:
$\tilde{G}^<_d = i\tilde{G}^r_d[\tilde{\Gamma}_L f_L
+\tilde{\Gamma}_R f_R ] \tilde{G}^a_d $, $\tilde{G}^>_d =
-i\tilde{G}^r_d \sum_{\alpha}
\tilde{\Gamma}_{\alpha}(1-f_{\alpha}) \tilde{G}^a_d $, and
$\tilde{G}^r_d(\omega) =\tilde{G}^{a*}_d(\omega) =1/\{\omega
-\tilde{\epsilon_d} +i\tilde{\Gamma}\}$, where $f_{\alpha}(\omega)
=1/\{exp[(\omega-\mu_{\alpha})/k_B T_e]+1\}$ is the Fermi
distribution, $\tilde{\Gamma}=(\tilde{\Gamma}_L+
\tilde{\Gamma}_R)/2$, and $\tilde{\Gamma}_{\alpha} =
\Gamma_{\alpha} exp[-(\lambda_q/\omega_q)^2(2N_q+1)]$ with the
line-width functions $\Gamma_{\alpha} \equiv 2\pi \rho_{\alpha}
|t_{\alpha}|^2$. Then after substituting these Green functions
into Eq.(12), and assuming that $\Gamma_{\alpha}$ is independent
of the energy $\omega$ (i.e. the wide linewidth
approximation),\cite{ref7} the heat generation is obtained as:
\begin{eqnarray}
  Q & =& \frac{2\omega_q \lambda_q^2}{\tilde{\Gamma}}
   \int \frac{d\omega}{2\pi} \left\{
    (N_e - N_q) \sum\limits_{\alpha} \tilde{\Gamma}_{\alpha}
    (\bar{f}_{\alpha}-f_{\alpha}) \right.\nonumber\\
   & + &\left.
    \frac{\tilde{\Gamma}_L\tilde{\Gamma}_R}{2\tilde{\Gamma}}
   (f_L-f_R)(\bar{f}_{L}-\bar{f}_{R}) \right\}
   {\bf Im} \tilde{G}_d^r(\omega)    {\bf Im}
   \tilde{G}_d^r(\bar{\omega}),
\end{eqnarray}
where $\bar{f}_{\alpha}=f_{\alpha}(\omega-\omega_q)$ and
$N_e=1/[exp(\omega_q/k_B T_e)-1]$. The heat generation of Eq.(13)
is separated into two terms. The first term is proportional to
$N_e -N_q$ and it represents the heat conduction between the
electron system and the phonon system due to the temperature
difference ($T_e \not= T_{ph}$). If $T_e > T_{ph}$, this term is
positive, meaning that the heat flows from the electron system to
the phonon system. On the other hand, if $T_e < T_{ph}$, this term
is negative. The second term describes the heat generation due to
the electronic current flowing through the device. In the case of
wide linewidth, this term is always positive regardless of the
values of $\mu_L$, $\mu_R$, $\epsilon_d$, and the temperatures
$T_e$ and $T_{ph}$. In other words, the heat is emitted when the
current flows through the QD. However, if $\Gamma_{\alpha}$
depends on the energy $\omega$, this term can be negative in some
cases. A negative $Q$ value means that the heat is absorbed when a
current is flowing, making this system work as a refrigerator.

We now numerically investigate the heat generation $Q$ for
symmetric barriers ($\tilde{\Gamma}_L =\tilde{\Gamma}_R =
\tilde{\Gamma}$) and with the same electron and phonon
temperature, $T_e=T_{ph} \equiv T$. Fig.1b shows $Q$ versus the
bias $V$ ($V=\mu_L -\mu_R$) at different temperatures. At zero
temperature ($T=0$), there exists a threshold bias $V_t
=\omega_q$, and the heat generation $Q$ vanishes when $V<V_t$,
independent of other parameters. The reason is that the process of
emitting a phonon requires an occupied state at $\omega$ and an
empty state at $\omega-\omega_q$, and this process is impossible
for $V<\omega_q$ at $T=0$. For non-zero temperatures, the heat
generation is still quite small so long as $V <V_t$. However, with
the bias increasing and passing $\omega_q +2\tilde{\Gamma}$, the
heat generation $Q$ increases rapidly. At a large bias limit, the
heat generation has a saturated value $\omega_q \lambda_q^2
\tilde{\Gamma}/(\omega_q^2 +4 \tilde{\Gamma}^2)$. The curve of $d
Q/ dV$ versus $V$ does not exhibit the phonon side-peaks. For a
comparison, the current $I$ versus the bias $V$ is also shown in
Fig.1b.\cite{note3} The current has no threshold bias and it is
non-zero even for very small bias $V$ at $T=0$. Moreover, the
curve of $dI/dV$-$V$ exhibits the phonon
side-peaks.\cite{ref4,ref6}

Figures 1c and 1d show the heat generation $Q$ and the current $I$
versus the QD's level $\tilde{\epsilon}_d$ at $T=0$ and for fixed
bias $V$. For $V<V_t$, $Q$ is zero everywhere, but the curve
$I$-$\tilde{\epsilon}_d$ has a peak at $\tilde{\epsilon}_d =0$.
When $V$ is slightly larger than $V_t=\omega_q$, the curve
$Q$-$\tilde{\epsilon}_d$ has two peaks at $\pm\omega_q/2$, while
the current $I$ is flat for $\tilde{\epsilon}_d$ within
$\pm\omega_q/2$. For a large bias $V$, $Q$ is large when
$\tilde{\epsilon}_d$ is around $0$. But the shape of the curve
$Q$-$\tilde{\epsilon}_d$ is different from that of
$I$-$\tilde{\epsilon}_d$. This means that the heat generation $Q$
is not proportional to $I$. However, it is well known that in a
macroscopic system, the Joule heat is proportional to $I$ for a
fixed bias $V$.

{\sl (ii) The ac bias case}. When the operator $\hat{X}$ in the
Hamiltonian (11) is replaced by $<\hat{X}>$, the electron-phonon
interaction is decoupled, and the Green functions, then the
time-dependent heat generation $Q(t)$ and the current $I(t)$, can
be easily obtained. Fig.2 shows the heat generation $Q(t)$ and the
current $I(t)$. The current $I(t)$ is a periodic function with the
period $2\pi/\omega_{ac}$, and the time-averaged current $<I(t)>$
is zero. The time-dependent heat generation $Q(t)$ is also
periodic, but with a different period $\pi/\omega_{ac}$, and the
time-averaged heat generation $<Q(t)>$ is always positive. In
particular, $Q(t)$ is non-zero even at the time $t$ when $I(t)=0$.

In summary, a general formula for the heat generation due to a
current flowing in nano-devices has been derived. This formula
expresses the heat transfer to the phonon system in terms of
two-particle non-equilibrium Green functions. Employing this
formula, we have investigated the heat generation in a
lead-QD-lead system. With a dc bias, there exists a threshold bias
$V_t$ at zero temperature. For a fixed bias $V$, the heat
generation is not proportional to the current $I$. Under an ac
bias, the time-dependent heat generation $Q(t)$ is studied and
$Q(t)$ is not zero even when the current $I(t)$ is zero. These
results are absent in macroscopic systems, unique to
nano-structures.

{\bf Acknowledgments:} We gratefully acknowledge the financial
support from the Chinese Academy of Sciences and NSF-China under
Grant Nos. 10474125 and 10525418. XCX is supported by US-DOE under
Grant No. DE-FG02-04ER46124 and NSF under CCF-052473.

\newpage

\begin{figure}

\caption{(Color online) (a) Schematic diagram for the system of a
central scattering region coupled to the leads and the phonon
reservoir. (b) The heat generation $Q$ (thick curves) and the
current $I$ (thin solid curve) vs. the bias $V$ at different
temperature $T$ with $\tilde{\epsilon}_d =0$. (c) and (d) are $Q$
and $I$ vs $\tilde{\epsilon}_d$ at $T=0$. The other parameters are
$\omega_q=\lambda_q=3$ and $\tilde{\Gamma}=1$. }\label{fig:1}

\caption{ The time-dependent heat generation $Q(t)$ (solid curve)
and the current $I$ (dotted thick curve) vs. the time $t$ with the
parameters $\omega_{ac} =5$, $\Delta = 10$,
$\tilde{\epsilon}_d=0.2$, $\omega_q=\lambda_q=3$,
$\tilde{\Gamma}=1$, and $T=0$. The thin dotted curve schematically
shows the ac-biased voltage $2\Delta \cos \omega_{ac} t$
}\label{fig:2}

\end{figure}


\begin{references}
\bibitem{note1}
In fact, even when the phonon distribution $N_q$ is
non-equilibrium, the general formulas (6) and (8) of the heat
generation still hold.
\bibitem{note2}
Note that the other term $\langle T_c [ \hat{d}_i^{\dagger}(\tau)
\hat{d}_j(\tau)]\rangle \langle T_c [\hat{d}_k^{\dagger}(0)
\hat{d}_l(0)] \rangle $ does not contribute to the heat
generation, since its Fourier transformation is zero when
$\omega_q \not= 0$.
\bibitem{note3}
In the current calculation, we use the same method as in
Ref.\cite{ref6}.
\end{references}
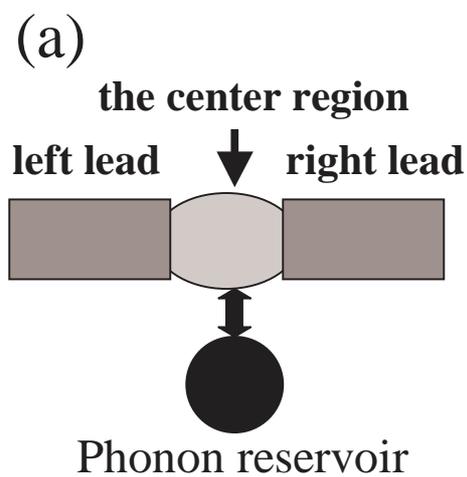
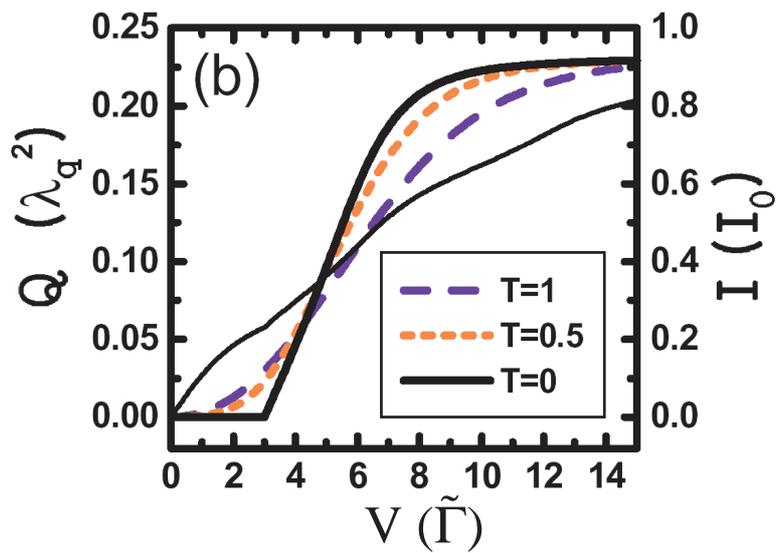
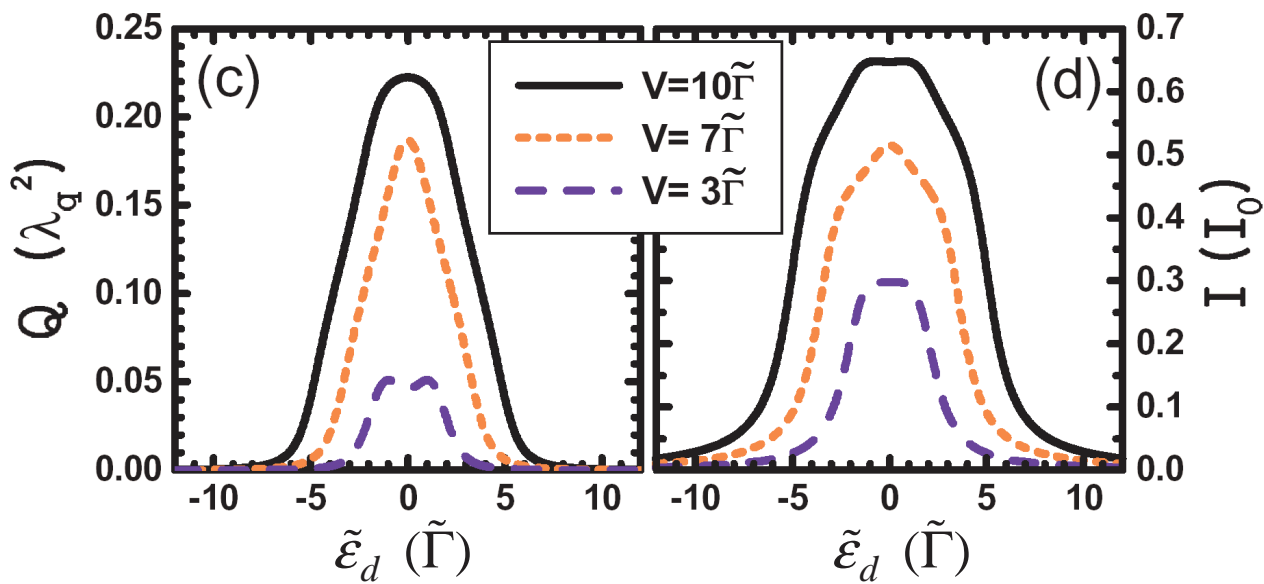
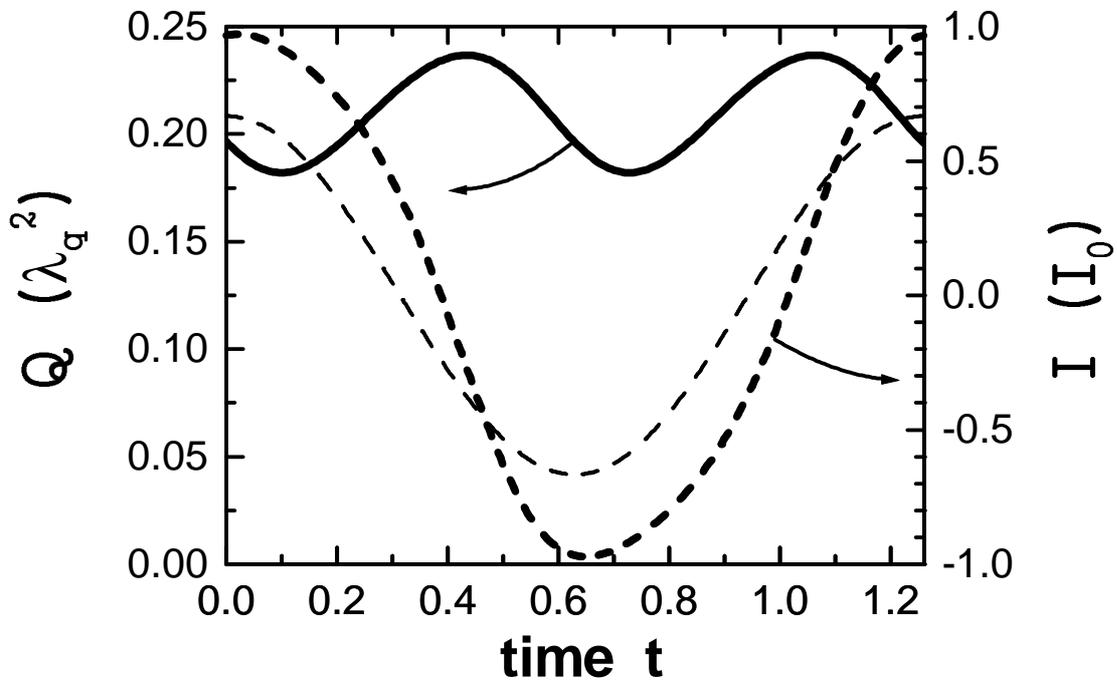
\end{document}